# GLOBAL COGNITIVE RADIO BASED COMMUNICATION SYSTEMS: SPACE-TIME COMMUNICATIONS

Dr. G. Rama Murthy[1] *rammurthy@iiit.ac.in*

M. Srikanth[2]  
*srikanth.munjuluri@research.iiit.ac.in*

K. Viswanadh[2]  
*konjeti.viswanadh.@research.iiit.ac.in*

## 1. INTRODUCTION:

Electro magnetic wave spectrum is limited. The available spectrum is divided into licensed and unlicensed bands. The demand for spectrum is increasing. Hence the need for innovation of new ideas for efficient spectrum utilization is increasingly felt. Also researchers are proposing utilization of ultra high frequency bands for electrical communication.

In an effort to find ways of meeting the increasing demand for spectrum, researchers tried observing the actual utilization (in licensed band) over the temporal and spatial dimensions. Surprisingly FCC found that actual utilization over certain time and spatial location was as low as 15%. Capitalizing on this observation, Joseph Mitola III proposed the idea of cognitive radio for dynamic spectrum access [1]. In section 2, details of cognitive radio as "currently understood" are explained.

The authors took a careful look at the spectrum utilization on a "global scale" (not just local utilization). They realized that Einstein's idea (logical basis of the theory of relativity) that space and time are not independent variables is highly relevant. For instance, when the time in Hyderabad is 12 noon, the time in New York is 2 AM. In this research paper, we explore capitalizing upon the idea of efficient spectrum utilization on a global scale (not just locally as proposed by J Mitola).

The paper is organised as follows: Section II overviews the conventional local cognitive radio technology. The "core" idea behind the paper is presented in section III. The concept of "Globally Cognitive Radio based Communication Systems" (GCRCS) is discussed in detail. In section IV, two practically viable scenarios are exemplified to illustrate the potential of this concept. The implementational issues and challenges are discussed in section V, followed by conclusion.

## 2. LOCAL COGNITIVE RADIO:

In the context of this paper, the term "local" is used to signify a particular geographical spatial location on Earth. A "local cognitive radio" refers to the conventional cognitive Radio (CR) technology, where, the cognition is applied on the temporal and spatial aspects "locally". A conventional CR focuses on the unused spectrum in a local region and attempts at the maximum utilisation of spectrum resources. Based on the methodology employed, spectrum access can be classified into two ways:

- Opportunistic Spectrum Access: The opportunistic spectrum access is the technique in which the secondary user will be allocated the spectrum in the guaranteed absence of primary user. This technique is employed when a primary user is absent for a substantial amount of time.
- Dynamic Spectrum Access: The dynamic spectrum access refers to more or less, the "hand-over" approach. A secondary user senses the spectrum continuously, and whenever a "spectrum hole" is sensed, the spectrum is utilised by the secondary user and again when the primary user is detected, the spectrum is handed over back to primary user. This method is employed when the presence of a primary user is probabilistic in nature.

The latter is more sophisticated than the former. The challenges involved in opportunistic and dynamic spectrum access techniques are studied in [2], [3].The basic tasks of cognitive radio include spectrum sensing, spectrum management, spectrum mobility and spectrum sharing. The challenges involved in each of the tasks are well understood [4],[5].CR technology finds its applications in many fields, especially in the subject of Wireless Sensor Networks [6]. Various enhancements have been proposed and soon, the effect of cognitive radio principles will be felt in daily life.

## 3. GLOBAL COGNITIVE RADIO:

### 3.1 Need for global cognition in Spectrum Access:

The previous section, local cognitive radio, would have brought a great relief to the former U.S. Federal Communications Commission (FCC) chair William Kennard, who coined the word, "spectrum drought", but there are still many reasons to worry about. The growing pace of multimedia applications is alarmingly high and hence it continues to pose a great challenge to wireless system designers. The methodology of local cognitive radio may not meet this ever increasing demand. So the danger of "spectrum drought" is not completely averted.

In this research paper, we propose the concept of global cognitive radio, where in, the problem of spectrum scarcity is countered with the efficient utilization of frequency resources on the global basis, not confining to a particular geographical region.

## 3.2 Principle of Global cognitive radio: (GCR)

The principle of global cognitive radio is the "**Global Opportunistic Remote Spectrum Access".** This principle is based on the observation that there is a highly deterministic trend in terms of spectrum usage in different countries at different times**.** This definite behaviour that can be observed in the spectrum usage at one particular place can be exploited from a distant place**.**

In this paper, the notions of up time and down times are used.

- Down Time is defined as the time at a spatial location, when radio transmissions (using the electromagnetic waves) by the licensed users are totally stopped.
- Up Time is defined as the time at a spatial location, when radio transmissions by licensed users are totally utilizing the spectrum to the maximum potential.

These Up Time and Down Time can occur over various diametrically opposite spatial locations on the earth. So the opportunistic spectrum access can be implemented on global scale. In addition, the Downtimes at various locations on planet keep shifting from one place to another, depicted in the below figure. Various other factors like the influence of variations of day and night times (among the different geographical regions) on up and down times are also to be studied. All these factors can be maximally utilised in the next sub section. The architecture of a Global Cognitive Radio System (GCRS) is discussed in the next sub section.

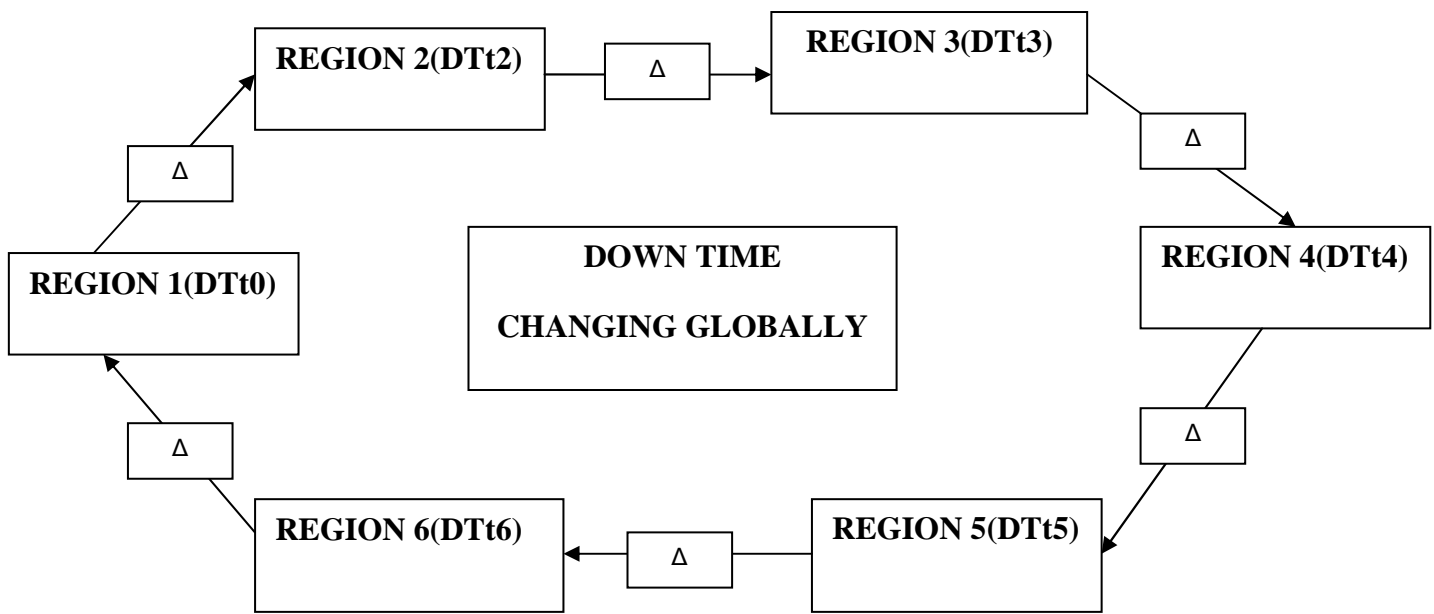

**DTtx – Downtime at time t = tx on Coordinated Universal Time (UTC)**

**Δ - Time delay**

### 3.3 Global Cognitive Radio System : (GCRS)

The Global Opportunistic Remote Spectrum Access can be implemented using the following architecture:

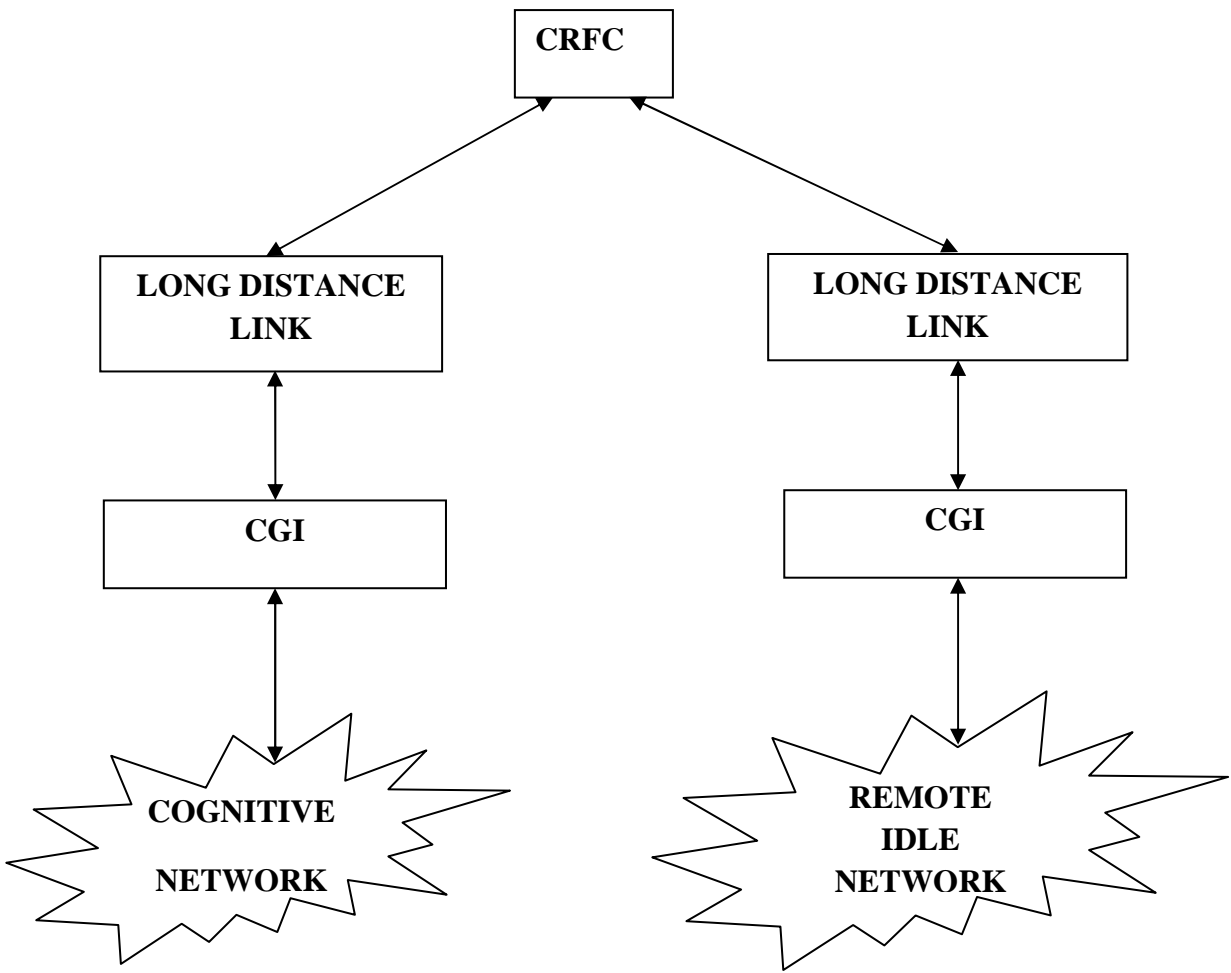

Explanation:

The architecture comprises of the following blocks:

- Cognitive Network: This network attempts to use the remotely idle spectrum.
- Remote Idle Network: This network acts as a spectrum resource that can be maximally utilised based on its availability.
- Common Gateway Interface (CGI): A communication gateway between the long distance link and networks (cognitive or remote).
- Long Distance Link: A link between CGI and CRFC.
- Cognitive Radio Function Coordinator (CRFC): The co-ordinator of the entire cognitive activity.

The cognitive and remote networks are connected to Common Gateway Interface (CGI). This is a crucial design block, which handles the complex responsibility of interfacing the network with the long distance links. The long distance links can be terrestrial or satellite links. The Coordinator is the heart of the architecture. It maintains the Universal Communication lookup Table (UCLT), which consists of the records of all the networks in the world, each of the records contains the information regarding the occupied and idle bands of the network, accessibility permissions and the duration of accessibility. During the busy time at "Cognitive Network" region, meaning, all its bands are full with primary users, the cognitive network would send a request to CRFC for enquiring about the availability of any vacant bands. CRFC then checks the UCLT and sends the required spectrum information.

## 4. ILLUSTRATIONS:

The concept of Global Cognitive Radio Architecture can be employed in various scenarios. Two crucial applications are illustrated in this section.

**4.1: Simple Opportunistic Remote Spectrum Access:**

A country in one specific time zone accesses the spectrum gaps of another country which is in some other time zone with significant time difference. So in other words, a country rents its idle spectrum to distant countries. This can be put to great use, especially in today's scenario of "flat-world" (Thomas L Friedman). For instance, as the Indian origin people are increasing in America, India can broadcast a Hindi FM service in America, accessing its unused bands (atleast for sometime).This would also help significantly in reducing the Internet bandwidth saving the Internet space from IP radio.

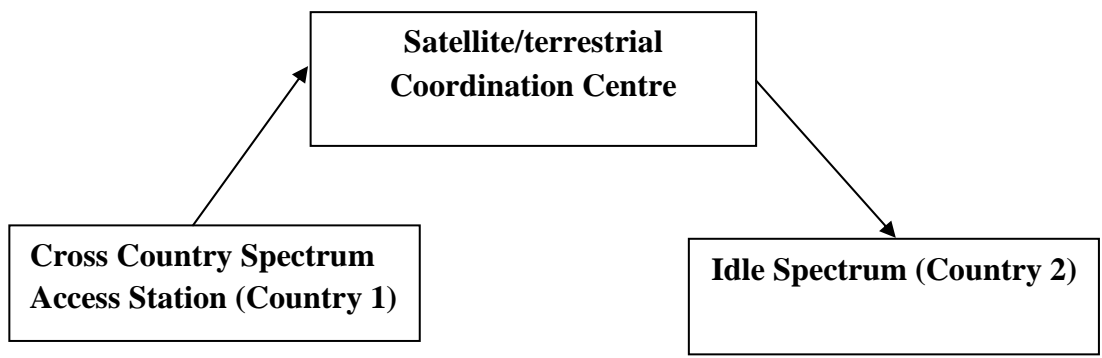

**4.2: Global Cognitive Radio with Inter-Satellite Communications:**

The concept of Global Cognitive Radio can be implemented using the constellation of satellites. The architecture of such a implantation is discussed below. We propose this scenario based on the following assumptions:

i) The entire Earth is globally connected in satellite space, i.e., the constellation of satellites (LEO or GEO) cover the entire Earth and every place on earth is accessible. It is reasonable to assume this from [7] and [8].

ii) Satellite and terrestrial communications can coexist in same frequency band. The concepts presented in [9] and [10] showed that the coexistence between satellite and terrestrial communications is possible with minimal interference.

The proposed scenario is elucidated in the below figure:

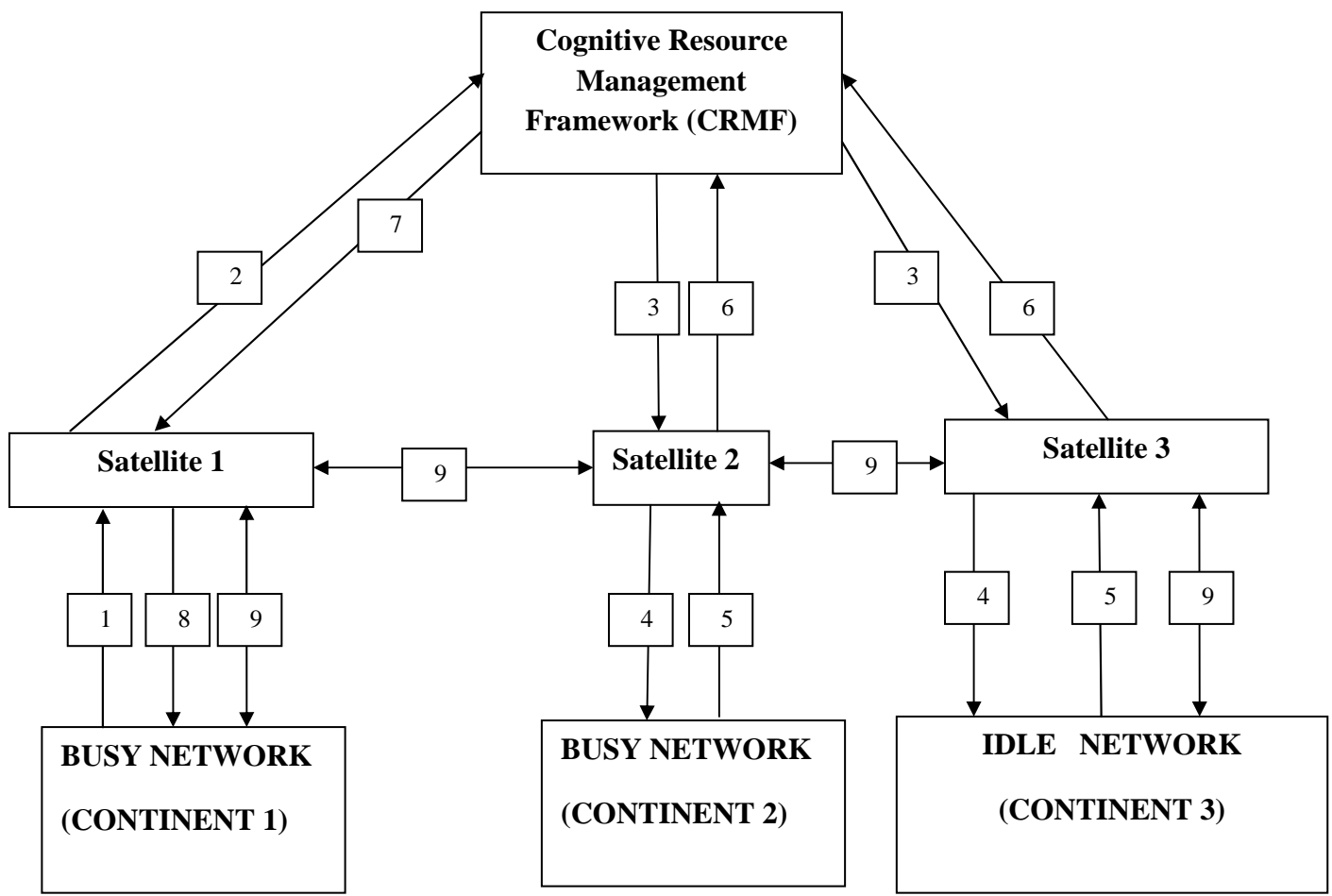

Explanation:

The above block diagram is explained below:

1) Busy network queries for any presence of idle spectrum.

2) Satellite forward the request to CRMF

3) CRMF sends a query to all its members (satellites) asking for the information of any spectrum

holes in their respective regions.

4) Satellites probe for any spectrum holes in their regions.

5) The regional network sends a idle or busy information back to satellite depending upon spectrum is idle or occupied respectively.

6) Satellites forward the status messages to CRMF.

7) CRMF forwards the status messages to the satellite, which requested for idle spectrum.

8) Satellite forwards the info to the respective regional network.

9) This is the most complex step of all the above. Once all the others parameters are fine, the cognitive link is established employing multiple hand over mechanism. This means the satellite(satellite 3), under whose jurisdiction a idle network is, hands over some of its spectrum responsibilities to the idle network and takes up the duties of the adjacent satellite(satellite 2).The adjacent satellite (satellite 2), hands over its duties to free satellite(satellite 3), and takes up the duties of its next satellite (satellite 1) and this goes on and ultimately the busy satellite(satellite 1) becomes free and this satellite can be used for communication purposes in the busy region(continent 1).This is possible because of our second assumption.

## 5. IMPLEMENTATION: Issues and Challenges:

In this section, we will discuss some of the issues and challenges relating to the practical implementation of the Global Cognitive Radio Concept.

**5.1: Challenges in Simple Opportunistic Remote Spectrum Access:**

When a country tries to access the "spectrum space" of another, there are many sorts of issues that are needed to be sort out.

Technical Issues:

1) The cross country spectrum access station can be designed in any of the three configurations; network-centric, distributed (ad hoc), and mesh architectures. The design challenges in all architectures are well studied in literature. But, so far, the access stations/base stations are equipped with local spectrum coordination and control, so collaborating with global coordination centres may introduce some new problems.

2) With the advent of various technologies in all spheres of science, there are multiple ways to

carry out point-to-point communications on Earth. Very far distance broadcasting can be implemented only using the satellite communications. But with the advent of Internet has revolutionised the things. IPTV and IP radio stand as very efficient and capable replacements for satellite communications. Thus the Coordination Centre can be implemented in either the satellite way or the Internet way. Satellite Communications based coordination centres are well studied. But using Internet to coordinate the spectrum space is relatively a new idea and has to be explored further.

Legal Issues:

As the spectrum access is across countries, issues like "Spectrum hacking" can take place. Regulatory bodies like ITU have to implement a robust framework for a fair and beneficial spectrum access across countries.

### 5.2: Global Cognitive Radio in Inter-Satellite Communications-Challenges:

This scenario is quite complex to implement as it involves the communications between many different protocols. It calls for an exchange of information and various hand-over mechanisms between terrestrial-terrestrial base stations, terrestrial-satellite base stations and satellite-satellite base stations. Hence the challenges in Satellite-Terrestrial Systems, Inter-Satellite communications and Terrestrial-terrestrial communications have to be well studied [11],[12].

# 6. Conclusion:

"Spectrum Scarcity" is a global problem across the world. This paper tried emphasising on the fact that a global problem has to be dealt on global basis, not just locally. The "Future Internet" and "Internet of Things" literally scare the communication system designer regarding the available bandwidth and spectrum resources. There is absolutely no scope to waste or underutilize the available resources. Hence the proposed idea of "GLOBAL COGNITIVE RADIO CONCEPT" can undoubtedly solve the resource problems in next Generation Communications.

## REFERENCES:

[1] J. Mitola, III, BCognitive radio,[Licentiatethesis], KTH, Royal Inst. of Technology,Stockholm, Sweden, Sep. 1999.

[2] C. Santivanez , R. Ramanathan , C. Partridge , R. Krishnan , M. Condell , S. Polit, Opportunistic spectrum access: challenges, architecture, protocols, Proceedings of the 2nd annual international

workshop on Wireless internet, p.13-es, August 02-05, 2006, Boston, Massachusetts.

[3] Milind M. Buddhikot , Alcatel-Lucent Bell Labs,Understanding Dynamic Spectrum Access: Models,Taxonomy and Challenges , April 17-21, 2007,Proceedings of IEEE DySPAN 2007, Dublin.

[4] Yucek, T.; Arslan, H., "A survey of spectrum sensing algorithms for cognitive radio applications," *Communications Surveys & Tutorials, IEEE* , vol.11, no.1, pp.116,130, First Quarter 2009.

[5] Akyildiz, I.F.; Won-Yeol Lee; Vuran, Mehmet C.; Mohanty, S., "A survey on spectrum management in cognitive radio networks," *Communications Magazine, IEEE* , vol.46, no.4, pp.40,48, April 2008.

[6] Kumar, Sumit, Deepti Singhal, and Rama Murthy Garimella. "Doubly Cognitive Architecture Based Cognitive Wireless Sensor Network." *arXiv preprint arXiv:1104.0142* (2011).

[7] Beste, D. C., "Design of Satellite Constellations for Optimal Continuous Coverage," *Aerospace and Electronic Systems, IEEE Transactions on* , vol.AES-14, no.3, pp.466,473, May 1978.

[8]Kandeepan, S.; De Nardis, L.; Di Benedetto, M.; Guidotti, A.; Corazza, G.E., "Cognitive Satellite Terrestrial Radios," *Global Telecommunications Conference (GLOBECOM 2010), 2010 IEEE* , vol., no., pp.1,6, 6-10 Dec. 2010.

[9]Deslandes, V.; Tronc, J.; Beylot, A.-L., "Analysis of interference issues in Integrated Satellite and Terrestrial Mobile Systems," *Advanced satellite multimedia systems conference (asma) and the 11th signal processing for space communications workshop (spsc), 2010 5th* , vol., no., pp.256,261, 13-15 Sept. 2010.

[10] http://www.itu.int/ITU-R/terrestrial/docs/fixedmobile/fxm-coord-shared.pdf.

[11] Taleb, T.; Hadjadj-Aoul, Y.; Ahmed, T., "Challenges, opportunities, and solutions for converged satellite and terrestrial networks," *Wireless Communications, IEEE* , vol.18, no.1, pp.46,52, February 2011.

[12] Evans, B.; Werner, M.; Lutz, E.; Bousquet, M.; Corazza, G.E.; Maral, G.; Rumeau, R., "Integration of satellite and terrestrial systems in future multimedia communications," *Wireless Communications, IEEE* , vol.12, no.5, pp.72,80, Oct. 2005.